\RequirePackage{fix-cm}
\documentclass[smallextended]{svjour3}       
\smartqed  
\usepackage{graphicx}
\usepackage{cite}
\usepackage{amsmath}
\def\msun{{\rm M_{\odot}}}

\def\be{\begin{equation}}
\def\ee{\end{equation}}

\def\del#1{{}}
\def\ltsima{$\; \buildrel < \over \sim \;$}
\def\simlt{\lower.5ex\hbox{\ltsima}}
\def\gtsima{$\; \buildrel > \over \sim \;$}
\def\simgt{\lower.5ex\hbox{\gtsima}}

\newcommand\aj{\rmfamily{AJ}, }%
\newcommand\araa{\rmfamily{ARA\&A}, }%
\newcommand\apj{\rmfamily{ApJ}, }%
\newcommand\apjl{\rmfamily{ApJ}, }%
\newcommand\aap{\rmfamily{A\&A}, }%
\newcommand\mnras{\rmfamily{MNRAS}, }%
\newcommand\nat{\rmfamily{Nature}, }%
\begin{document}

\title{The $M-\sigma$ relation between supermassive black holes and their host galaxies
}


\author{Kastytis Zubovas         \and
        Andrew R. King 
}


\institute{K. Zubovas \at
              Center for Physical Sciences and Technology, Saul\.{e}tekio al. 3, Vilnius, Lithuania \\
              Vilnius University Observatory, Saul\.{e}tekio al. 3, Vilnius, Lithuania \\
              Tel.: +370-601-20485\\
              \email{kastytis.zubovas@ftmc.lt}           
           \and
           A. R. King \at
              Department of Physics and Astronomy, University of Leicester, University Road, LE1 7RH Leicester, United Kingdom \\
              Astronomical Institute Anton Pannekoek, University of Amsterdam, Science Park 904, 1098 XH Amsterdam, Netherlands\\ 
              Leiden Observatory, Leiden University, Niels Bohrweg 2, NL-2333 CA Leiden, Netherlands \\
}

\date{Received: date / Accepted: date}

\maketitle

\begin{abstract}
Supermassive black holes (SMBHs) are found in the centres of most galaxies. Their masses, and hence their gravitational potentials, are negligible compared with those of the host galaxy. However, several strong correlations between SMBH masses and host galaxy properties have been observed, notably the $M-\sigma$ relation connecting the SMBH mass to the characteristic velocity of stars in the galaxy. The existence of these correlations implies that the SMBH influences the evolution of its host galaxy. In this review, we present the most promising physical model of this influence, known as the Active galactic nucleus (AGN) wind feedback model. Winds launched from the accretion disc around the SMBH can drive powerful outflows, provided that the SMBH is massive enough - this condition establishes the $M-\sigma$ relation. Outflows can have a profound influence on the evolution of the host galaxy, by compressing its gas and driving it out, affecting the star formation rate. We present the current status of the model and the observational evidence for it, as well as the directions of future research.
\end{abstract}

\section{Introduction}\label{sec:intro}

The existence of supermassive black holes (SMBHs) in the nuclei of most galaxies was first proposed in the 1960s by Salpeter \cite{Salpeter1964ApJ} and Lynden-Bell \cite{Lynden-Bell1969Natur}. The basic argument for their existence is the following. Some galaxies are observed to have very bright nuclei, radiating as much energy as the rest of the galaxy's stars put together. Nuclear fusion, which powers stars, converts only $0.7\%$ of matter into radiation, and therefore would require unfeasible amounts of hydrogen gas fusing in galactic nuclei to explain the observed luminosities. Accretion of matter on to a black hole, on the other hand, can liberate between $5.5\%$ and $\sim 42\%$ of the rest mass energy of the infalling material, with the average value of $\sim 10\%$. This would lead to the brightest nuclei consuming $<20 \msun {\rm yr}^{-1}$ of matter, where $\msun = 1.989 \times 10^{33}$~g is the mass of the Sun. While this is still a formidable amount, it is certainly achievable given the amounts of gas available close to galactic centres. 

Soon afterward, development of the accretion disc theory \cite{Shakura1973A&A} suggested a way to detect accreting black holes. Accretion discs around SMBHs would produce significant amounts of electromagnetic radiation, mainly in the ultraviolet part of the spectrum, which would contribute to the background radiation budget of the Universe. Calculation of the amount of radiation produced by accretion of a given amount of material and comparison of this with the observed radiation levels allowed Soltan \cite{Soltan1982MNRAS} to calculate that, on average, every large galaxy should contain an SMBH that grew by accreting gas. These black holes must then reside mainly in the centres of their galaxies, since their gravitational interactions with the surrounding stars produce wakes that dampen the SMBH kinetic energy, causing them to ``sink'' down to the bottom of the potential well.

More direct evidence of the existence of SMBHs in galactic centres came about in the last decade of the 20th century. It included the detection of gravitational redshift in spectral lines of gas very close to the nucleus of the active galaxy MCG-6-30-15 \cite{Tanaka1995Natur}, as well as high-resolution observations of stellar and gas kinematics \cite{Harms1994ApJ, Kormendy1995ARA&A}. Furthermore, observations of our Galactic centre revealed that the radio source Sgr A* \cite{Balick1974ApJ} is powered by an SMBH \cite{Falcke1993A&A, Krichbaum1993A&A, Schodel2002Natur}.

Systematic searches for SMBHs, undertaken at the turn of the century \cite{Magorrian1998AJ}, revealed a number of surprising correlations between black hole masses and host galaxy properties. The most interesting one of them is the correlation with $\sigma$, the velocity dispersion of the stars in the spheroidal part of the galaxy (i.e. everything other than the disc) \cite{Ferrarese2000ApJ}, although many others exist. The interpretation of these correlations has led to the development of the field of active galactic nucleus (AGN) feedback on the host galaxy, and the co-evolution between SMBHs and their host galaxies. Discovery of small-scale quasi-relativistic winds \cite{Pounds2003MNRASa, Pounds2003MNRASb, Tombesi2010A&A, Tombesi2010ApJ} and large-scale massive outflows \cite{Feruglio2010A&A, Rupke2011ApJ, Sturm2011ApJ, Cicone2014A&A} in AGN has further strengthened the argument that AGN feedback is an important element of galaxy evolution.

In this review, we present the observed correlations and their significance (Section \ref{sec:obs}). We then discuss the proposed explanations for the origin of these correlations (Section \ref{sec:expl}) and describe in detail the wind feedback model, which is one of the most successful in explaining the observed data (Section \ref{sec:wind_fb}). Finally, we outline some of the prospects for the near future in Section \ref{sec:prospects} and conclude in Section \ref{sec:concl}.

\section{The observed correlations} \label{sec:obs}

A black hole in general relativity has only three properties: mass, angular momentum and electric charge. In astrophysical contexts, electric charge is generally unimportant, since it oscillates around zero (a positively charged black hole predominantly attracts electrons, leading to a charge decrease, and vice versa). Measuring angular momentum of real black holes is extremely difficult, with only a few results obtained so far \cite{Risaliti2013Natur,Valtonen2016ApJ}. Therefore, the observed correlations involve only the mass of the SMBH, $M_{\rm BH}$.

The galaxy properties that correlate with SMBH mass are surprisingly diverse, from the mass \cite{Haering2004ApJ} and velocity dispersion \cite{Ferrarese2000ApJ} of the galaxy spheroid, to the total mass of globular clusters in the galaxy \cite{Burkert2010ApJ} or the gravitational binding energy of the galaxy bulge \cite{Aller2007ApJ}. Of these numerous correlations, the most fundamental one appears to be the correlation between SMBH mass and the host galaxy spheroid velocity dispersion, the $M-\sigma$ relation \cite{Ferrarese2000ApJ,Gultekin2009ApJ,McConnell2013ApJ}.

\subsection{Measuring the relevant parameters}

Several methods can be used to determine the mass of a SMBH, all of them based on simple Newtonian dynamics. Relativistic effects, such as precession and frame dragging, are only relevant on spatial scales of several thousand Schwarzschild radii $R_{\rm S}$ and lower, but the gravitational influence of a SMBH extends out to a distance (cf. \cite{Peebles1972ApJ})
\begin{equation}\label{eq:rinf}
    R_{\rm inf} = \frac{2GM_{\rm BH}}{\sigma^2} \simeq 20 M_8 \sigma_{200}^{-2} {\rm pc} = 2.25 \times 10^6 \sigma_{200}^{-2} R_{\rm S}.
\end{equation}
Here, $\sigma \equiv 200 \sigma_{200}$~km~s$^{-1}$ is the velocity dispersion in the host galaxy, and the SMBH mass was parameterized to $M_{\rm BH} \equiv 10^8 M_8 \, \msun$. Gas and stellar motions are usually measured well within the sphere of influence, but well outside the region where relativistic effects become noticeable, therefore Newtonian dynamics is a reasonable approximation to the actual motion. Velocities are typically measured using spectroscopy and utilising Doppler shifts of approaching/receding material, while distances can be calculated by measuring the light travel time from the AGN itself to the nearby gas clouds. Once this has been done, the SMBH mass is calculated by 
\begin{equation}
    M_{\rm BH} = f\frac{R^2v}{G},
\end{equation}
where $f$ is a factor of order a few that encompasses the deviations from a circular orbit.

The galaxy parameter most important for us is the velocity dispersion $\sigma$, which is the characteristic velocity with which stars move in the spheroidal part of the galaxy - the bulge of a spiral galaxy or an elliptical galaxy as a whole. It is also measured by spectroscopic observations of the galaxy. Knowing the velocity dispersion and the size of the galaxy, one can calculate the mass of the spheroid, from which further parameters can be derived and correlations with $M_{\rm BH}$ investigated.

\subsection{The $M-\sigma$ relation} \label{sec:msigma}

\begin{figure*}
  \includegraphics[width=\textwidth]{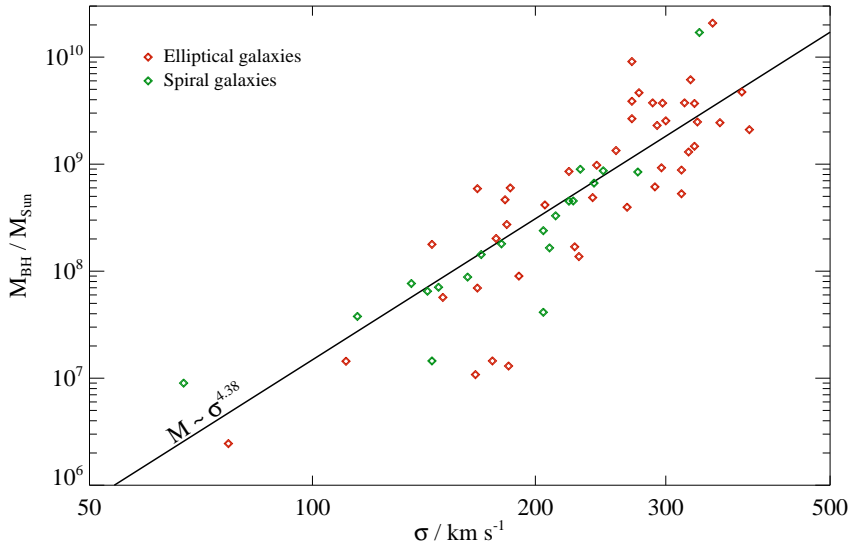}
\caption{Values of velocity dispersion $\sigma$ and central black hole mass $M_{\rm BH}$ for a sample of elliptical (red points) and spiral (green points) galaxies. Data taken from \cite{Kormendy2013ARA&A}, figure produced by authors. The data shows a power-law correlation $M \propto \sigma^{4.38}$, shown as a straight line.}
\label{fig:msigma}
\end{figure*}

The $M-\sigma$ relation connects the mass of the SMBH with the velocity dispersion in the galaxy spheroid. It was first discovered in 2000 \cite{Ferrarese2000ApJ} and is given by a power law:
\begin{equation}
    \frac{M_{\rm BH}}{10^8 \; \msun} \simeq A \left(\frac{\sigma}{200 {\rm km s}^{-1}}\right)^\alpha,
\end{equation}
with the scaling values chosen to correspond to typical SMBH masses and velocity dispersions in galaxies, and the parameters having values $A \sim 3$ and $\alpha \sim 4-6$. The precise value of $\alpha$ depends on the sample of galaxies investigated; different authors suggest values as varied as $\alpha \simeq 4.24\pm0.41$ \cite{Gultekin2009ApJ}, $\alpha \simeq 4.8\pm0.5$ \cite{Ferrarese2000ApJ} and $\alpha \simeq 5.64$ \cite{McConnell2013ApJ}. In Figure \ref{fig:msigma}, we reproduce the data collected in \cite{Kormendy2013ARA&A}, which gives $\alpha \simeq 4.38$. The relation has a scatter of $\sim0.4$~dex \cite{McConnell2013ApJ}.

It is possible that the observed relation is merely an upper limit. In order to determine the mass of the black hole, its sphere of influence has to be resolved, therefore at any given value of $\sigma$, there may be many undermassive black holes that are not detected \cite{Batcheldor2010ApJ}; however, this seems unlikely to be the case \cite{Gultekin2011ApJ}. Furthermore, this requirement forces the two parameters to be at least somewhat correlated, since the sphere of influence depends on $\sigma$ (see eq. \ref{eq:rinf}), possibly making the observed correlation flatter than the intrinsic one \cite{Morabito2012ApJ}.

The existence of such a relation is remarkable. As mentioned above, the gravitational influence of the SMBH extends only to the inner several tens of parsecs of the galaxy (eq. \ref{eq:rinf}). Spheroids usually have sizes of a kiloparsec or more, so the gravity of the SMBH cannot control the rest of the galaxy. Yet somehow, the bulge, or even the whole spiral galaxy, seems to be aware of the presence of the SMBH and of its mass; or, alternatively, the growth of the SMBH is governed by the same processes that govern the growth of the galaxy.

\subsection{Correlations with other central massive objects} \label{sec:other_cmos}

Many galaxies host nuclear star clusters (NSCs) in addition to, or instead of, nuclear supermassive black holes \cite{Boker2010IAUS}. In smaller galaxies, those with velocity dispersion $\sigma \simlt 150$~km~s$^{-1}$, NSCs are the dominant central massive object \cite{Nayakshin2009MNRAS}. 

The mass of the NSC, similarly to that of the SMBH, correlates with the galaxy velocity dispersion, although differently. There are two, seemingly equally possible, proposed correlations: one with the same slope as the $M-\sigma$ relation, but with a different scaling, such that $M_{\rm NSC}/M_{\rm BH} \simeq 20$ at a given value of $\sigma$ \cite{Ferrarese2006ApJ}; and one with a different slope: $M_{\rm NSC} \propto \sigma^{2.11\pm0.31}$ \cite{Scott2013ApJ}. Additionally, the mass of the NSC correlates with the mass of the SMBH, if any is present: $M_{\rm BH} \propto M_{\rm NSC}^{2.7\pm0.7}$ \cite{Graham2016IAUS}.

The existence of these correlations suggests that similar processes control the interaction between the galaxy and its central massive object, independently of whether the object is a SMBH or a NSC. The details of those processes may differ somewhat, leading to the differences in observed correlations.

\section{The proposed explanations} \label{sec:expl}

Any explanation of the observed correlation must account for the vastly different physical scales of the two objects. The Schwarzschild radius of the SMBH is
\begin{equation}
    R_{\rm S} = \frac{2GM_{\rm BH}}{c^2} \simeq 10^{-5} M_8 {\rm pc}.
\end{equation}
The sphere of influence also extends only to several tens of parsecs at most (eq. \ref{eq:rinf}). Both values are much smaller than typical sizes of galaxy bulges ($R_{\rm b} \sim 1$~kpc) and dark matter haloes ($R_{\rm h,vir} \sim 200$~kpc). There is no possibility that the gravity of the SMBH controls the motions of stars in the bulge.

Over the years, a number of models have been proposed attempting to explain the origin of $M-\sigma$ and other observed correlation. They can be loosely grouped into three categories:
\begin{itemize}
    \item ``Central limit theorem'' explanations, asserting there is no causal connection, but since both galaxies and black holes grow via mergers, their mass ratio approaches a constant value as a galaxy evolves \cite{Peng2007ApJ,Jahnke2011ApJ}. In this case, the $M-\sigma$ relation appears due to the fact that a galaxy's mass is roughly proportional to $\sigma^4$ (the Faber-Jackson relation \cite{Faber1976ApJ}).
    \item ``Feeding rate'' explanations, which are based on the assumption that the galaxy controls the rate at which gas is fed to the SMBH. This rate should depend on gas velocities in the galactic spheroid, and therefore may lead to the SMBH growing to a limiting mass set by the properties of the galaxy \cite{Haan2009ApJ,Angles-Alcazar2013ApJ,Angles2015ApJ}.
    \item ``Feedback'' explanations, based on the energy release by matter accretion on to the SMBH, which may affect the whole galaxy and even larger intergalactic scales. There are several forms of feedback, which we present below, some of which are definitely capable of controlling star formation throughout the host galaxy of the SMBH. Furthermore, feedback explanations can explain many of the observed properties of active galaxies, such as the presence of massive outflows \cite{Fiore2017A&A} and fast nuclear winds \cite{Tombesi2013MNRAS}.
\end{itemize}

\subsection{Black hole feedback} \label{sec:feedback}

The most promising possibility of connecting SMBHs with their host galaxies, feedback is a general term for any kind of process whereby matter falling on to a SMBH releases energy that, in turn, affects subsequent matter infall. It is easy to see that energy release during the build-up of a black hole can be significant:
\begin{equation} \label{eq:ebh}
    E_{\rm BH} \simeq \eta M_{\rm BH} c^2 \simeq 2\times10^{61} M_8 \eta_{0.1} \;{\rm erg},
\end{equation}
where $\eta \sim 0.1$ is the radiative efficiency of accretion. This value is much higher than the binding energy of a bulge likely to host this black hole:
\begin{equation} \label{eq:ebulge}
    E_{\rm b} \simeq M_{\rm b} \sigma^2 \simeq 8\times10^{58} M_{11} \sigma_{200}^2 \;{\rm erg}.
\end{equation}
A coupling efficiency of only $f_{\rm E} \simeq 8\times 10^{58} / 2\times 10^{61} \simeq 4\times10^{-3}$ would be enough to unbind the galaxy bulge. Removing gas from the bulge requires even less energy, by a factor $f_{\rm g} \equiv M_{\rm gas}/M_{\rm b}$. The question then becomes not whether AGN can affect their host galaxies, but how come the energy coupling efficiency is so small that black holes actually grow to the observed masses without destroying their own, and the galaxy's, gas supply. Feedback can manifest via several mechanisms, or modes, that satisfy this constraint and explain some of the observed phenomena.

A major difference between feedback modes is the accretion rate, and therefore luminosity, of the AGN. It is typically expressed as a ratio between the actual luminosity and the Eddington luminosity, a theoretical limit at which radiation pressure force on the surrounding plasma is capable of counteracting the SMBH gravity. Eddington luminosity is directly proportional to the SMBH mass:
\begin{equation}
    L_{\rm Edd} = \frac{4 \pi G M_{\rm BH} c}{\kappa_{\rm e.s.}};
\end{equation}
here, $\kappa_{\rm e.s.} = 0.348$~cm$^{2}$~g$^{-1}$ is the electron scattering opacity.

AGN with luminosities $L < 0.01 L_{\rm Edd}$ produce jets - powerful streams of matter moving at relativistic velocities - as the main form of power output \cite{Falcke2004A&A,Merloni2007MNRAS}. Jets mainly affect the environment on scales of galaxy clusters where energy injection offsets the cooling of gas halos \cite{McNamara2007ARA&A}, preventing gas accretion on to galaxies and thus regulating the star formation rate. This type of feedback can explain the galaxy luminosity function and the ages of stars in elliptical galaxies \cite{Croton2006MNRAS,DeLucia2007MNRAS,Sijacki2007MNRAS,Marulli2008MNRAS}. This so-called ``maintenance mode'' feedback is the primary form of feedback for the majority of each AGN's lifetime \cite{Jones2016ApJ}.

Radiatively efficient AGN, i.e. those with $L > 0.01 L_{\rm Edd}$, heat some gas in the galactic nucleus \cite{Sazonov2005MNRAS}, but their main effect on the surroundings comes from radiation pressure and winds. Dusty interstellar medium (ISM) gas has a much higher opacity to radiation than the electron scattering opacity $\kappa_{\rm e.s.}$, therefore it can be pushed away by an AGN with $L < L_{\rm Edd}$. The effective Eddington limit for dusty gas can be as much as 500 times lower than that for ionized gas \cite{Fabian2008MNRAS}. This may lead to AGN radiation pushing gas out of the bulge and limiting its growth \cite{Ishibashi2012MNRAS}. On the other hand, radiation can evaporate the dust grains that produce opacity, allowing radiation to escape from the galaxy \cite{Barnes2018arXiv}. Therefore, radiation pressure effects are probably important only in nuclear regions rather than throughout the whole bulge \cite{Costa2018MNRAS}.

More efficient coupling between AGN luminosity and the surrounding gas can be achieved via winds, which are launched by the AGN radiation field from the accretion disc around the SMBH and then push against the surrounding ISM. If the wind carries and transfers a constant fraction of the AGN luminosity to the surrounding gas, one can show that the gas is pushed away on a condition $L_{\rm AGN} > L_{\rm crit} \propto \sigma^5$ \cite{Silk1998A&A}. Equating the AGN luminosity to a constant fraction of its Eddington luminosity leads to a relation $M_{\rm BH} \propto \sigma^5$, which has a slope very similar to that of the observed relation. However, the offset of the predicted relation is much lower than observed, unless various arbitrary assumptions are made regarding the porosity of the ISM, the coupling efficiency and so on \cite{Silk1998A&A,King2010MNRASb}. On the other hand, if a significant fraction of the ISM is pushed only by the wind momentum, the resulting relation becomes $M_{\rm BH} \propto \sigma^4$, with a slope slightly shallower than observed, but the offset corresponding very closely to the observed black hole masses, at least at $\sigma \simeq 200$~km~s$^{-1}$ \cite{King2003ApJ,King2010MNRASa}. The wind energy can be either radiated away or used to drive a powerful large-scale outflow, affecting the whole galaxy bulge \cite{Zubovas2012ApJ}. In the next Section, we present this wind feedback model in more detail.



\section{The wind feedback model} \label{sec:wind_fb}

In this section, we present the physical basis and the main results of the wind feedback model. First proposed in 2003 \cite{King2003ApJ} and based on then-recent detection of ultra-fast outflows in AGN \cite{Pounds2003MNRASa,Pounds2003MNRASb}, the model has since been developed to explain a large number of observed phenomena on scales from the sub-parsec environment of SMBHs to the circumgalactic medium. 

\subsection{AGN winds} \label{sec:wind}

The primary basis of the feedback mechanism is the launching of AGN winds from the accretion disc. Radiation from the accretion disc and the corona surrounding the AGN pushes some of the gas away from the disc, forming a quasi-spherical wind. The wind momentum is given by
\begin{equation}
    \dot{M}_{\rm w} v_{\rm w} = \tau \frac{L_{\rm AGN}}{c};
\end{equation}
here, the quantities on the left-hand side are the mass flow rate and velocity of the wind, and on the right $L_{\rm AGN}$ is the AGN luminosity and $c$ is the speed of light. The quantity $\tau$ is the optical depth of the wind - the average number of times a photon emitted by the AGN is scattered by the wind material before escaping to infinity. Since the luminosity comes from accretion, we may write $L_{\rm AGN} = \eta \dot{M}_{\rm BH} c^2$ and express the wind velocity as
\begin{equation}
    v_{\rm w} = \frac{\tau}{\dot{m}} \eta c = 0.1 \frac{\tau \eta_{0.1}}{\dot{m}} c,
\end{equation}
where $\dot{m} \equiv \dot{M}_{\rm w}/\dot{M}_{\rm BH}$ is the ratio of the wind mass flow rate to SMBH accretion rate. Observed winds, usually called Ultra-fast outflows (UFOs), tend to have $\tau \simeq 1$ \cite{King2003MNRASb} and similarly, $\dot{m}\sim 1$ \cite{King2010MNRASa,King2010MNRASb}. The small variations of these parameters around unity explain the range of observed UFO velocities $0.03 c < v_{\rm w} < 0.3 c$ \cite{Tombesi2010A&A, Tombesi2010ApJ}. The UFOs are detected in a significant fraction, $\sim 40\%$, of all nearby AGN \cite{Tombesi2010A&A, Tombesi2010ApJ}, implying that these features appear frequently or almost continuously and have wide covering angles (i.e. are quasi-spherical).

The launch radius $R_{\rm l}$ of the wind should be approximately the distance at which the circular velocity $v_{\rm circ} = v_{\rm w}$, i.e. of order $R_{\rm l} \sim 100 R_{\rm S} \sim 10^{15} M_8$~cm. Typical distances at which UFO are observed, $R_{\rm UFO} = 10^{14-18}$~cm \cite{Tombesi2012MNRAS} then imply a travel time of
\begin{equation}
    t_{\rm w} \simeq \frac{R_{\rm UFO} - R_{\rm l}}{v_{\rm w}} < 10 {\rm yr};
\end{equation}
this timescale is of order months for most UFOs, implying that AGN winds are a very intermittent phenomenon.

The gas comprising the wind is very strongly ionized. This is typically quantified using an ionization parameter $\xi$, which reflects the relative density of ionizing photons and gas particles in the medium. It is defined as
\begin{equation}
    \xi \equiv \frac{L_{\rm ion}}{nR^2} \equiv \frac{l_{\rm ion}L_{\rm AGN} \mu m_{\rm p}}{\rho R^2},
\end{equation}
where $L_{\rm ion} \equiv l_{\rm ion} L_{\rm AGN}$ is the ionizing luminosity, $\mu = 0.63$ is the atomic weight per particle for fully ionized material and $m_{\rm p} = 1.67 \times 10^{24}$~g is the proton mass. A value of $\xi \sim 40-100$ corresponds to one ionizing photon per atom (the actual value depends on the precise spectrum of the AGN radiation). For AGN winds, we find
\begin{equation}
    \xi = 4 \pi b l_{\rm ion} \mu m_{\rm p}  \frac{\eta^2 \tau}{\dot{m}^2} c^3 \simeq 3.6 \times 10^4 \frac{l_{\rm ion}}{0.01} \frac{\eta_{0.1}^2 \tau}{\dot{m}^2} {\rm erg}\, {\rm cm} \, {\rm s}^{-1},
\end{equation}
i.e. the winds are extremely strongly ionized. This value is consistent with observational data.

The kinetic energy rate carried by the wind is
\begin{equation}
    \dot{E}_{\rm w} = \frac{\dot{M}_{\rm w} v_{\rm w}^2}{2} = \frac{\tau^2 \eta^2}{2 \dot{m}} \dot{M}_{\rm BH} c^2 = \frac{\tau^2 \eta}{2 \dot{m}} L_{\rm AGN} = 0.05 \frac{\tau^2 \eta_{0.1}}{\dot{m}} L_{\rm AGN}.
\end{equation}
The wind therefore carries a large fraction of the energy liberated by luminous accretion. This energy is more than enough to unbind the bulge of the galaxy (see equations \ref{eq:ebh} and \ref{eq:ebulge}), provided that it is efficiently communicated to the ISM at larger scales.

\subsection{Wind shock and cooling} \label{sec:wind_cooling}

The wind velocity is significantly larger than the velocity dispersion of the host galaxy spheroid, $\sigma \sim 200$~km~s$^{-1}$. Therefore, once the wind reaches the surrounding ISM, a strong shock develops. The shock temperature is
\begin{equation}
    T_{\rm sh} = \frac{3 \mu m_{\rm p} v_{\rm w}^2}{16 k} \simeq 1.3 \times 10^{10} \frac{\tau^2 \eta^2_{0.1}}{\dot{m}^2} {\rm K} \gg T_{\rm w}.
\end{equation}
The wind heats up and slows down, and most of its original kinetic energy is converted into thermal. In effect, AGN wind fuels an extremely hot small bubble in the centre of the galaxy. Its subsequent evolution is determined by two factors: the distance of the shock from the AGN and the geometric structure of the ISM.

\begin{figure*}
  \includegraphics[width=\textwidth]{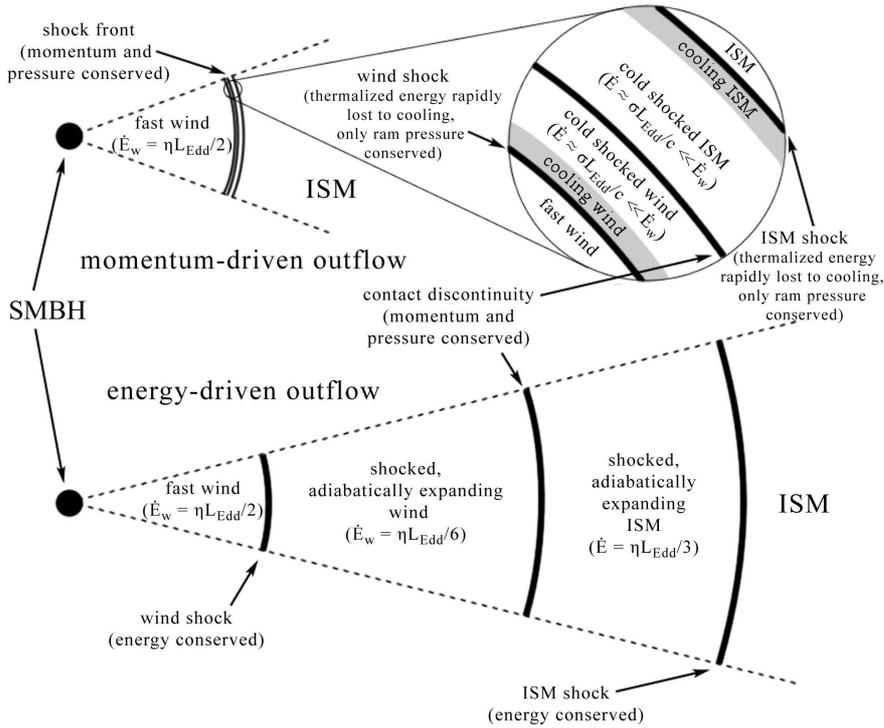}
\caption{Schematic illustration of the two feedback modes: momentum-driven outflow (top) and energy-driven outflow (bottom). Figure taken from \cite{Zubovas2012ApJ}.}
\label{fig:outflows}
\end{figure*}

The distance of the shock front determines how efficiently the shocked wind cools. The only efficient cooling mechanism is the inverse Compton effect: interaction between photons emitted by the AGN and electrons hotter than the Compton temperature $T_{\rm C} \sim 2\times 10^7$~K. The density of the radiation field decreases as $R^{-2}$, therefore the cooling time increases as $R^2$. The bubble cannot expand faster than at its own sound speed, therefore the constant-velocity expansion timescale increases as $R$. At some distance $R_{\rm cool}$ from the AGN, cooling becomes inefficient and the bubble becomes effectively adiabatic. This transition delineates two distinct modes of feedback: outflows forming within $R_{\rm cool}$ are driven primarily by the momentum of the AGN radiation field, while those outside $R_{\rm cool}$ are driven by the total wind energy. The schematic view of the two types of outflow is presented in Figure \ref{fig:outflows}, taken from \cite{Zubovas2012ApJ}.

The ISM geometry can complicate this picture, however. As an adiabatic bubble expands, it encounters regions of different density in different directions. Therefore, a large fraction of the energy contained in the bubble can escape through low-density channels in the ISM, with the high density material (such as molecular clouds) effectively feeling only the momentum of the AGN radiation field \cite{Zubovas2014MNRASb}. These `leftover' clouds can continue feeding the SMBH even while a large-scale outflow is ongoing in the galaxy. Feeding only ceases once the wind momentum becomes large enough to push away the dense gas. We show below that it is this momentum condition that establishes the $M-\sigma$ relation.

The value of $R_{\rm cool}$ is particularly important when analysing the behaviour of outflows. If we treat the shocked wind as a one-temperature plasma, inverse-Compton cooling is very efficient out to $R_{\rm cool} \simeq 500$~pc for typical galaxy parameters \cite{King2003ApJ}. In this case, momentum-driven outflows occur in the centre of the galaxy and only transition to energy-driven ones when pushed far out from the galaxy's centre. However, two-temperature effects might be important \cite{Faucher2012MNRASb}. The electrons in the shocked wind cool efficiently, but most of the energy is contained in the protons. If the electron-proton equilibration timescale governs the temperature evolution, then cooling is only efficient within $R_{\rm cool} < 1$~pc from the nucleus. In this case, essentially all outflows are adiabatic, but the wind momentum is still important for removing the dense gas. We shall show below that the momentum-driven outflows are important for establishing the $M-\sigma$ relation, but the large-scale energy-driven outflows can affect the whole host galaxy.

\subsection{Momentum-driven outflows} \label{sec:mom_outflow}

Dense gas, or gas close to the SMBH, is pushed outward mainly by the wind momentum. The equation of motion for this gas is just the appropriate expression of Newton's second law:
\begin{equation}\label{eq:eom_mom}
    \frac{{\rm d}}{{\rm d}t}\left[M\left(R\right) \dot{R}\right] = \tau \frac{L_{\rm AGN}}{c} - \frac{G M\left(R\right) M_{\rm b} \left(R\right)}{R^2}.
\end{equation}
Here, the left-hand term is the change in momentum of the outflowing material, which, at a radial distance $R$, contains gas with mass $M\left(R\right)$ and is moving at a velocity $\dot{R}$. The first term on the right-hand side is the driving force, and the second term is the force of gravity, with $M_{\rm b}\left(R\right)$ the background mass contained within $R$, including stars, dark matter and the SMBH. If we consider the case where $R \gg R_{\rm inf}$, the SMBH gravity is not important. Now if we further simplify the problem by assuming that all matter is distributed isothermally,
\begin{equation}
    \rho_{\rm tot}\left(R\right) = \frac{\sigma^2}{2\pi G R^2}; \; M_{\rm tot}\left(R\right) = \frac{2 \sigma^2}{G} R,
\end{equation}
we can rewrite eq. (\ref{eq:eom_mom}) as
\begin{equation}
    \frac{2f_{\rm g}\sigma^2}{G} \frac{{\rm d}}{{\rm d}t}\left(R \dot{R}\right) = \tau \frac{L_{\rm AGN}}{c} - \frac{4f_{\rm g} \left(1-f_{\rm g}\right)\sigma^4}{G},
\end{equation}
where $f_{\rm g} \equiv \rho_{\rm g}/\rho_{\rm tot}$. If we multiply both sides by $R \dot{R}$ and rearrange slightly, we end up with
\begin{equation}
    \frac{{\rm d}}{{\rm d}t}\left[\left(R \dot{R}\right)^2\right] = \frac{G}{2f_{\rm g}\sigma^2} \left[\tau \frac{L_{\rm AGN}}{c} - \frac{4f_{\rm g} \left(1-f_{\rm g}\right)\sigma^4}{G}\right] \frac{{\rm d}}{{\rm d}t}\left(R^2\right).
\end{equation}
Integrating both sides with respect to time, we end up with (cf. \cite{King2005ApJ,King2010MNRASa})
\begin{equation}
    \dot{R}^2 = \frac{G}{2f_{\rm g}\sigma^2} \left[\tau \frac{L_{\rm AGN}}{c} - \frac{4f_{\rm g} \left(1-f_{\rm g}\right)\sigma^4}{G}\right] + \frac{C}{R^2},
\end{equation}
where $C$ is the integration constant. At large radii, the $C/R^2$ term becomes negligible and the shell can only expand indefinitely if the term in the square brackets is positive. Rearranging, we find a critical AGN luminosity condition (cf. \cite{King2003ApJ,Murray2005ApJ,King2010MNRASa}):
\begin{equation}
    L_{\rm AGN} > L_{\rm crit} = \frac{4 f_{\rm g} \left(1-f_{\rm g}\right) \sigma^4 c}{G \tau}.
\end{equation}
Taking $\tau \simeq 1$ and $L_{\rm AGN} = l L_{\rm Edd} = 4 \pi G M_{\rm BH} c l / \kappa_{\rm e.s.}$, we can convert this into a critical mass condition for the maximum luminosity $l=1$:
\begin{equation}
    M_{\rm BH} > M_{\sigma} = \frac{\kappa_{\rm e.s.} f_{\rm g} \left(1-f_{\rm g}\right) \sigma^4}{\pi G^2} \simeq 2.7\times 10^8 \sigma_{200}^4 \, \msun.
\end{equation}
In the last equality, we used $f_{\rm g} = 0.16$, which is the ratio between ordinary matter and total matter density in the Universe. The derived relation has a slope similar to, if slightly shallower than, the observed $M-\sigma$ relation, and the intercept is only a factor $\sim2-3$ higher than typical observed values at $\sigma = 200$~km~s$^{-1}$. Considering that no free parameters are used in the derivation, this agreement is remarkable.

Analytical solutions for the evolution of outflow properties are impossible for more realistic background density profiles. However, assuming spherical symmetry, the equation of motion can be integrated numerically for any arbitrary mass distribution. This reveals essentially similar evolution, with a critical mass $M_{\rm crit}\simeq M_{\sigma}$ required for the outflow to reach large distances \cite{McQuillin2013MNRAS}. Similarly, outflow evolution is almost the same when the central engine is a nuclear star cluster rather than an AGN, although the critical mass is larger by a factor $\sim 20$ or higher \cite{McLaughlin2006ApJ,Bourne2016MNRAS}. The difference is partly due to the lower radiative efficiency of thermonuclear fusion and partly due to the cancellation of winds ejected by individual stars.

Outflows may occur in galaxies long before $M_{\rm BH}$ reaches the critical $M_\sigma$ value, but they cannot remove dense gas. Material is pushed slightly outward, then stalls and falls back. Such churning motions might explain one important feature of the central parts of galactic dark matter halos: the fact that they tend to have $\rho_{\rm DM} \simeq const.$ in the central regions \cite{deBlok2001ApJ,Gentile2004MNRAS,Gilmore2007ApJ}. Pure-dark-matter numerical simulations predict cuspy profiles with inner slopes $\rho_{\rm DM} \propto R^{-1}$ \cite{Navarro1997ApJ}, so it is likely that gas motions affect the dark matter density profile. In dwarf galaxies, repeated gas expulsion by supernova feedback flattens the central density peaks  \cite{Governato2012MNRAS,Pontzen2012MNRAS}. In larger galaxies, the same effect may be achieved by repeated bursts of AGN feedback, with outflows reaching several hundred parsec distances before collapsing.

\subsection{Energy-driven outflows} \label{sec:en_outflow}
  
On large scales, and for diffuse gas, shocked wind no longer cools efficiently and so most of the kinetic energy of the wind is transferred to the surrounding gas. In this case, we can understand the evolution of the outflow by considering both the equation of motion and the energy equation. The former in this case reads (cf. \cite{King2005ApJ,Zubovas2016MNRASb})
\begin{equation}\label{eq:eom_en}
    \frac{{\rm d}}{{\rm d}t}\left[M\left(R\right) \dot{R}\right] = 4 \pi R^2 P - \frac{4 \pi G M\left(R\right) M_{\rm tot} \left(R\right)}{R^2}.
\end{equation}
The terms in this equation have the same meaning as in equation \ref{eq:eom_mom}, and $P$ is the pressure of the shocked wind bubble. The latter is constrained by the energy equation:
\begin{equation}\label{eq:energy}
    \frac{{\rm d}}{{\rm d}t}\left[\frac{3}{2}PV\right] = \frac{\eta}{2}L_{\rm AGN} - P\frac{{\rm d}V}{{\rm d}t} - \frac{{\rm d}E_{\rm g}}{{\rm d}t},
\end{equation}
where $V$ is the volume of the outflow bubble and $E_{\rm g}$ is the gravitational binding energy. Using eq. \ref{eq:eom_en} to eliminate $P$ in eq. \ref{eq:energy} leads, after some algebra, to a numerically-integrable equation of motion:
\begin{equation}
\begin{split}
    \dddot{R} &= \frac{\eta L_{\rm AGN}}{M R} - \frac{2\dot{M} \ddot{R}}{M} - \frac{3\dot{M} \dot{R}^2}{M R} - \frac{3\dot{R} \ddot{R}}{R} - \frac{\ddot{M} \dot{R}}{M} \\ & +\frac{G}{R^2}\left[\dot{M} + \dot{M}_{\rm b} + \dot{M}\frac{M_{\rm b}}{M} - \frac{3}{2}\left(2M_{\rm b}+M\right)\frac{\dot{R}}{R}\right].
\end{split}
\end{equation}
Here, the time derivatives of the mass are defined as $\dot{M}\equiv \dot{R}\partial M/\partial R$ and $\ddot{M} \equiv \ddot{R}\partial M/\partial R + \dot{R} ({\rm d}/{\rm d}t)\left(\partial M/\partial R\right)$. For details of the derivation, we refer the reader to \cite{Zubovas2016MNRASb}.

Much like in the case of momentum-driven outflow, some analytical insight can be gained by considering a much more simplified case of an isothermal matter distribution with gas fraction $f'_{\rm g}$, which may be lower than the cosmological value $f_{\rm c} = 0.16$. In this case, further assuming that $L_{\rm AGN} = lL_{\rm Edd}$ and $M_{\rm BH} = M_\sigma$, the equation of motion simplifies to (cf. \cite{King2011MNRAS})
\begin{equation}
    \eta l c \sigma^2 \frac{f_{\rm c}}{f'_{\rm g}} = 5 \sigma^2 \dot{R} + \frac{3}{2}\dot{R}^3 + 3 R \dot{R} \ddot{R} + \frac{1}{2}R^2 \dddot{R}.
\end{equation}
This equation has a solution of the form $R = v_{\rm e} t$, i.e. $\dot{R} = const.$; importantly, this solution is attractive, i.e. outflows tend to follow it independently of initial conditions. Substituting this expression into the equation above, we find
\begin{equation}
    2 \eta l c \frac{f_{\rm c}}{f'_{\rm g}} = 10 v_{\rm e} + 3\frac{v_{\rm e}^3}{\sigma^2}.
\end{equation}
This equation can be solved by assuming either $v_{\rm e} \gg \sigma$, which eliminates the first term on the right-hand side, or $v_{\rm e} \ll \sigma$, which eliminates the second. The second assumption leads to a contradiction, so the correct solution is
\begin{equation}
    v_{\rm e} \simeq \left(\frac{2 \eta l c \sigma^2 f_{\rm c}}{3 f'_{\rm g}}\right)^{1/3} \simeq 930 \left(\frac{\eta_{0.1}lf_{\rm c}}{f'_{\rm g}}\right)^{1/3} \sigma_{200}^{2/3} \, {\rm km}\, {\rm s}^{-1}.
\end{equation}
This velocity is much higher than typical galactic escape velocities $v_{\rm esc} \sim \sqrt{2}\sigma \sim 300$~km~s$^{-1}$, so energy-driven outflows can be easily pushed out of the galaxy entirely. These outflows carry the material enriched by stellar processes from the centre of the galaxy to the circumgalactic medium (CGM), enriching it significantly \cite{Oppenheimer2010MNRAS,Tumlinson2011Sci}.

Large-scale outflows have now been observed in many AGN (e.g. \cite{Feruglio2010A&A,Rupke2011ApJ,Sturm2011ApJ,Cicone2014A&A,Fiore2017A&A}) with properties generally in agreement with those predicted by this model \cite{Zubovas2012ApJ}. Furthermore, the simultaneous discovery of an AGN wind and a large-scale outflow in a single galaxy is considered strong evidence for the connection between the two features \cite{Tombesi2015Natur,Veilleux2017ApJ}.

It is, however, important to consider the timescales relevant for AGN and outflows. The duration of a single AGN episode is $t_{\rm AGN} \sim 5\times10^4 - 10^5$~yr \cite{Schawinski2015MNRAS,King2015MNRAS}. This is much shorter than the outflow dynamical timescale $t_{\rm out} \simgt 10^6$~yr, therefore it is highly unlikely that the AGN episode observed simultaneously with an outflow is the one responsible for inflating that outflow \cite{Zubovas2018MNRAS,Nardini2018MNRAS}. In fact, once the AGN switches off, the outflow continues expanding for approximately an order of magnitude longer time than the duration of the AGN episode \cite{King2011MNRAS}, therefore most observed large-scale outflows might be `fossil' outflows left over from earlier AGN episodes. This situation had both advantages and drawbacks: understanding the connection between an outflow and the AGN episode that inflated it is difficult, but it may also provide information about the activity history of the galaxy that would be impossible to obtain otherwise.

\subsection{Effects on star formation} \label{sec:starform}

Energy-driven outflows described in the previous section can have a profound influence on the evolution of the host galaxy. In particular, the mass outflow rate can easily reach $\dot{M}_{\rm out} > 10^3 \, \msun$~yr$^{-1} \gg \dot{M}_*$, where $\dot{M}_*$ is the star formation rate of the host galaxy \cite{Zubovas2012ApJ,Feruglio2010A&A}. Therefore, outflows can remove vast amounts of gas from the galaxy before that gas turns into stars, effectively sweeping the galaxy out and shutting down star formation. This process takes several tens of millions of years and is much faster than consumption of gas by star formation, which takes billions of years. Therefore energy-driven AGN outflows can rapidly transform galaxies from star-forming to quiescent; this result is consistent with most AGN being found in galaxies that are intermediate between the two regimes \cite{Schawinski2007MNRAS}.

While the outflow is moving through the galaxy, it can actually enhance the star formation rate. This happens in two ways: via cooling of the outflowing material and via compression of the galaxy disc.

Outflow cooling is a rapid process. The forward shock that the outflow drives into the ISM heats the gas to temperatures of order $10^7$~K. At this temperature, all molecules are easily destroyed by collisions, and the gas is fully ionized. However, the gas is also compressed, and therefore cools very rapidly, on timescales far shorter than the dynamical timescale \cite{Zubovas2014MNRASa,Richings2018MNRAS,Richings2018MNRASb}. This allows molecules to reform and explains the presence of massive amounts of molecular gas, which is the main observational evidence of these outflows. Compressed molecular gas is also gravitationally unstable, fragments and can easily form stars \cite{Nayakshin2012MNRASb,Zubovas2013MNRAS}. If we assume that star formation is controlled by the feedback from young stars (in the form of winds, photoionization and supernova explosions) balancing the rate of cooling, the star formation rate (SFR) is (cf. \cite{Zubovas2014MNRASa})
\begin{equation}
    \dot{M}_* \simeq 80 \epsilon_{\rm f}^{-1} R_{\rm kpc}^{-1}\left(\frac{f_{\rm g}}{f_{\rm c}}\right)^{2.08} \sigma_{200}^{10/3} \, \msun {\rm yr}^{-1},
\end{equation}
where $\epsilon_{\rm f}$ is the fraction of stellar energy release that couples to the gas and $R_{\rm kpc}$ is the outflow radius in kiloparsecs. This SFR is rather large, almost two orders of magnitude higher than the SFR of the Milky Way \cite{Robitaille2010ApJ,Licquia2015ApJ}. Recent observations revealed that star formation in outflows is a widespread phenomenon \cite{Maiolino2017Natur,Gallagher2018arXiv}.

In spiral galaxies, the disc material has too much weight to be pushed out by the outflow directly, therefore the outflow expands around the disc and compresses it \cite{Zubovas2013MNRASb}. This compression can be strong enough to enhance star formation rates in the disc by up to an order of magnitude \cite{Zubovas2016MNRASa}. The net result of this process is a rapid quenching of star formation in the galaxy disc as well as in the spheroid, since enhanced star formation consumes most of the disc gas and, once the outflow has passed and external compression is reduced, the SFR decreases to much lower values than before the burst.

\section{Prospects for the future} \label{sec:prospects}

The wind-driven outflow model is remarkably successful in explaining various observed features of AGN outflows. Several questions still remain, requiring further research and providing opportunities for testing the model.

Analytical arguments can only provide some insight into AGN outflow evolution, but the details and the interplay among various complex processes can only be probed numerically. Large-scale hydrodynamic simulations of galaxy evolution currently lack the resolution to model accretion on to the SMBH and AGN outflows self-consistently. The typical approach, injection of feedback energy into the gas surrounding the SMBH, actually produces unphysical results, resulting in outflows that are significantly too weak and small-scale \cite{Zubovas2016MNRASa}. Low-resolution simulations also fail to capture the whole range of effects of AGN outflows, resulting in them being too destructive \cite{Bourne2015MNRAS}. In fact, idealised high-resolution simulations show that AGN outflows are a complex phenomenon, quenching star formation in some regions of the galaxy spheroid, while enhancing it in others \cite{Zubovas2017MNRAS}. In the future, detailed galaxy evolution simulations must incorporate the physically-motivated models of AGN outflows, as well as appropriate numerical prescriptions for injecting AGN feedback energy and momentum, in order to reveal the full scope of AGN effects on their host galaxies.

Observations can help test some predictions of the model as it stands. For example, star formation in AGN outflows has been detected recently \cite{Maiolino2017Natur,Gallagher2018arXiv}, but other significant features of stars formed due to the action of outflows may be identified in the near future. Stars formed from outflowing material itself will have extremely radial orbits, perhaps even escaping their host galaxies \cite{Zubovas2013MNRAS}. Detecting such stars and quantifying their contribution to extragalactic light, as well as analysing supernova explosions outside galaxies, will immensely help constrain the magnitude of this positive feedback that AGN have on star formation. Similarly, discs experiencing starbursts due to outflow-induced compression will show a radial age gradient of young stars, corresponding to the timescale of outflow propagation. This gradient may be rather shallow, but in principle detectable as our understanding of stellar evolution improves.

Since the timescale of outflow evolution is longer than the typical AGN episode duration \cite{Schawinski2015MNRAS,King2015MNRAS}, a single galaxy may harbour several radially separated outflows driven by separate AGN episodes. This may be the case in, for example, IRAS F11119+3257 \cite{Nardini2018MNRAS}. Evolution of multiple outflows driven by a variable AGN is, obviously, much more complicated than the evolution of a single well-defined outflow. This problem may still be tractable in a semi-analytical fashion without recourse to hydrodynamic simulations, but that is not guaranteed so far. Ideally, understanding how multiple outflows evolve and interact would help pin down the properties of the galaxy and its nuclear activity history much more precisely than using a single outflow.

\section{Summary} \label{sec:concl}

Over the past twenty years, the astronomical community recognized the immense importance of supermassive black holes and AGN to the evolution of galaxies. Correlations between SMBH masses and host galaxy properties provided the first hints that a connection between these objects exists, while later developments, both observational and theoretical, clarified the nature and magnitude of this connection. AGN-driven outflows are now routinely being detected. Theoretical advances, such as the wind-driven outflow model presented in detail in this review, help interpret the data and understand how AGN affect their galaxies. Numerical simulations help us quantify how much AGN affect their host galaxies over cosmological timescales.

Outflows serve one more potentially important role, as dynamical footprints of past AGN activity. Persisting for several times $10^6$~yr after the AGN itself switches off, they encode information about both the AGN episode that inflated them and the galactic ISM they expanded in. Understanding how outflow properties depend on those of the AGN and the host galaxy will allow us to use outflows as probes of the past properties in their host galaxy. This should be immensely helpful when trying to understand the long-term properties of AGN variability, as well as the details of galactic, and especially ISM, evolution on megayear timescales.

\end{document}